\documentclass[12pt]{article}
\usepackage{graphicx}
\usepackage{amsmath}
\usepackage{amsfonts}
\usepackage{amssymb}
\newcommand{\be}{\begin{equation}}
\newcommand{\ee}{\end{equation}}
\newcommand{\bq}{\begin{eqnarray}}
\newcommand{\eq}{\end{eqnarray}}

\begin{document}

\title{\textbf{On the Meaning of the String-Inspired Noncommutativity and its
Implications}}
\author{\ \textbf{\ \ \ G. Dourado Barbosa} \\
%EndAName
\\
{\small \ \ Centro Brasileiro de Pesquisas F\'{i}sicas, CBPF-CCP}\\
{\small Rua Dr. Xavier Sigaud 150 \ , 22290-180, Rio de Janeiro, Brazil}}
\maketitle

\begin{abstract}
We propose an alternative interpretation for the meaning of noncommutativity
of the string-inspired field theories and quantum mechanics. Arguments are
presented to show that the noncommutativity generated in the stringy context
should\ be assumed to be only between the particle coordinate observables,
and not of the spacetime coordinates. Some implications of this fact for
noncomutative field theories and quantum mechanics are discussed. In
particular, a consistent interpretation is given for the wavefunction in
quantum mechanics.\ An analysis of the noncommutative theories\ in the
Schr\"{o}dinger formulation is performed employing a generalized quantum
Hamilton-Jacobi formalism. A formal structure for noncommutative quantum
mechanics, richer than the one of noncommutative quantum field theory, comes
out. Conditions for the classical and commutative limits of these theories
have also been determined and applied in some examples.\vspace{4cm}

email: gbarbosa@cbpf.br

Keywords: string theory, noncommutative field theory, Schr\"{o}dinger
formulation,

Hamilton-Jacobi formalism, quantum mechanics, classical and commutative
limits

PACS numbers :11.10.Lm; 11.10.Ef; 03.65.-w; 03.65.Bz
\end{abstract}

\section{Introduction}

Recently, there has been a great interest in noncommutative theories of
the\ canonical type. They are characterized by the following commutation
relation
\begin{equation}
\lbrack\widehat{X}^{\mu},\widehat{X}^{\nu}]=i\theta^{\mu\nu}\;,\label{1}%
\end{equation}
where $\theta^{\mu\nu}$ is an antisymmetric constant tensor function of the coordinates.

During the latest years, a great deal of work and effort has been done in the
direction of understanding this kind of noncommutativity and its implications
in the context of quantum\ field theory and quantum mechanics (for good
reviews see \cite{1,1.5}). Part of the interest in these theories is due to
the fact that noncommutativity is present in certain models of string theory
\cite{2,3}, and M-theory \cite{4,5}. Another motivation comes from field
theory itself. Semi-classical arguments, combining General Relativity and the
Heisenberg uncertainty principle, also lead to noncommutative field theory
\cite{6}. The study of these theories\ gives us the opportunity to understand
interesting phenomena, like nonlocality and IR/UV mixing \cite{7}, new physics
at very short distances \cite{1.5,8}, and possible implications of Lorentz
violation \cite{9}.

But the interest in noncommutativity goes beyond formal manipulations. There
are several lines of investigation on the possible phenomenology associated to
the canonical noncommutativity, from cosmology and high energy scattering
experiments to low energy precision tests\ \cite{10}.

The articles in this emerging branch of physics are growing up very fast. By
carefully analyzing the literature in the area, it is possible to identify two
primary research lines in canonical noncommutative quantum field theory. The
first one presupposes a noncommutative spacetime from the beginning,\ being an
intrinsic approach. It is closely related to the ideas of Connes
noncommutative geometry \cite{12}. This is the line followed, for example, by
Chaichian et all \cite{13}, Cho et all \cite{14} and\ Wulkenhaar \cite{15}.

The second research line has its roots in a low-energy approximation
of\ string theory in\ the presence of a strong background\ $B$\ field
\cite{3}. This is the approach assumed, for example, in \cite{7} and
followers. It is also the one adopted in this work. Therefore, we will refer
to the noncommutative quantum field theory (NCQFT) and\ the associated
noncommutative quantum mechanics (NCQM) always adopting this point of view.

The implications of considering noncommutativity as property of the spacetime
have been a subject under discussion. As it\ can be seen in \cite{7}, NCQFT is
supported by the use of the Feynman graph approach for its calculations. It
was argued in \cite{15}, for example, that such kind of procedure would not be
expected from a theory defined on a true noncommutative spacetime, which not
a\ manifold, at least in the usual sense. At this point, it is important to
take into account that\ NCQFT comes from string theory, which is based on a
continuous manifold structure. Therefore, it would be quite reasonable to
interpret\ it\ as commutative as long as spacetime is concerned. This
naturally leads to the formulation of a question: what kind of
noncommutativity can be behind NCQFT, if not the one of spacetime?

The first goal of this work is to present an answer to the question above,
proposing an alternative interpretation for the string-inspired
noncommutativity and discussing its implications for NCQFT and NCQM. As we
shall see later, one of these implications is the need of reinterpreting the
meaning of the wavefunction. Since the interpretation for the noncommutativity
proposed in this work is entirely new, we feel the necessity to analyze more
carefully the NCQFT and NCQM formulations until now available in the
literature. This is quite related to our second goal, which is to explore the
Schr\"{o}dinger formulation of NCQFT and NCQM, whose advantages to discuss
some fundamental questions, like the quantum/classical and
noncommutative/commutative passages, will be underlined in our exposition.

The organization of this work is the following. In Section 2, we give a new
interpretation for the meaning of the noncommutativity originated in the
string theory context by tracing an intuitive parallel with the Landau
problem. The Third Section deals with NCQFT in the Schr\"{o}dinger picture and
the discussion of the non-triviality to achieve its commutative and classical
limits. The remaining of the work is on NCQM. In the Fourth Section, we show
how to construct a consistent interpretation for NCQM in the Schr\"{o}dinger
formulation, a problem still now open in the literature. We also apply the
method proposed in Section 3 to carry out an analysis of the formal structure
of quantum mechanics and to determine the conditions for the achievement of
its classical and commutative limits. Two simple examples of application are
given in Section 5 to illustrate how the ideas proposed here work in concrete
models.\ We end up in Section 6 with a general discussion and outlook.

\section{ The Meaning of Stringy\ Noncommutativity}

Let us discuss the origin of the noncommutativity relation (\ref{1}) and its
interpretation. For simplicity, we will restrict our work to the case where
$\theta^{oi}=0$. To grasp some intuition, we will appeal for the similarity
between the derivation of the canonical noncommutativity in the stringy
context and the one present in the Landau problem when a system is projected
onto its lowest Landau level \cite{1,3}. Though being exhaustively explored in
the literature, this analogy has always been centered in what is the origin of
the noncommutativity,\ and not in what it is about.

We start by analyzing the Landau problem. Consider a nonrelativistic particle
moving on the $x-y$ plane in the presence of a constant magnetic$\ B $ field
pointing in the $z$-direction. The classical Lagrangian of the system is
\begin{equation}
L=\frac{1}{2}m\overset{\cdot}{\vec{x}}^{2}+e\overset{\cdot}{\vec{x}}\cdot
\vec{A}\text{ ,}\label{2}%
\end{equation}
where $e$ is the particle charge, $m$ is the particle mass, and $\vec{A}$ is
the electromagnetic vector potential. We are considering units where $c=1.$
The quantum Hamiltonian is
\begin{equation}
\widehat{H}=\frac{1}{2m}\hat{\pi}^{i}\hat{\pi}^{i},\label{2.5}%
\end{equation}
where $\hat{\pi}^{i}=m\overset{\cdot}{\hat{x}^{i}}=\hat{p}^{i}-e\hat{A}^{i}$
are the physical momenta and $\hat{p}^{i}$ are\ the canonical momenta. Notice
that the canonical momenta commute, while the physical momenta satisfy the
commutation relation
\begin{equation}
\lbrack\hat{\pi}^{i},\hat{\pi}^{j}]=i\hbar eB\epsilon^{ij}.\label{2.7}%
\end{equation}
To understand how the noncommutativity (\ref{1}) arises in the model\ it is
useful to define, in analogy with the classical case, the center-of-orbit
operator, whose components are given by
\begin{equation}
\widehat{X}^{i}=\hat{x}^{i}-\frac{i}{eB}\hat{\pi}^{i}.\label{2.8}%
\end{equation}
These components can be shown to satisfy the commutation relation
\begin{equation}
\lbrack\widehat{X}^{i},\widehat{X}^{j}]=-i\frac{\hbar\epsilon^{ij}}%
{eB}=i\theta^{ij},\label{2.9}%
\end{equation}
where $\theta^{ij}=\left(  -\hbar/eB\right)  \epsilon^{ij}$. Notice that while
$[\hat{x}^{i},\hat{x}^{j}]=0$, the $\widehat{X}^{i}$ are not allowed to
commute due to the presence of the term containing the magnetic field.\ The
spacetime, on the other hand, is the same ordinary commutative one. In this
scenario, the\ uncertainty relation
\begin{equation}
\Delta X^{i}\Delta X^{j}\geq\frac{\hbar}{2}\left|  \frac{\epsilon^{ij}}%
{eB}\right|  ,\label{2.95}%
\end{equation}
\ introduced by (\ref{2.9}), must be understood as just being a consequence of
the\ limitation on the information available about the $X^{i}$ coordinates\ in
a process of measurement.

Now we consider\ the strong magnetic field limit. In this case, the system is
projected onto the lowest Landau level. A rigorous prescription of how to work
in this limit, which is achieved by solving the constraint $\hat{\pi}%
^{i}\approx0$\ (using a projection technique), may be found in \cite{15.3}. On
heuristic grounds, one can understand the projection onto the lowest Landau
level as a process where the particles have their kinetic degrees of freedom
frozen and are confined into their respective orbit centers \cite{15.4}. The
particle coordinate observables in this limit clearly satisfy (\ref{2.9}) as a
consequence of the coincidence between $\widehat{X}^{i}$ and $\hat{x}^{i}$,
but there is no fundamental reason to believe that the spacetime coordinates
will become noncommutative in this limit as a consequence of the projection.
In reality the spacetime should be assumed as being unaltered, instead of
becoming fuzzy or pointless.

Usually the relation (\ref{2.9}) is achieved in the literature by dropping the
kinetic term directly from the Lagrangian (\ref{2}).\ If we write the vector
potential as $\overrightarrow{A}=\left(  0,Bx,0\right)  ,$ and consider the
$B\rightarrow\infty$ or $m\rightarrow0$ limits, we can discard the kinetic
term and write the Lagrangian as
\begin{equation}
L=eBx\dot{y}.\label{3}%
\end{equation}
In this Lagrangian, the $x$ and $y$ variables are canonically conjugate, and
their respective quantum operators satisfy a commutation relation identical
to\ (\ref{2.9}),
\begin{equation}
\left[  \hat{x}^{i},\hat{x}^{j}\right]  =-\frac{i\hbar}{eB}\epsilon
^{ij}.\label{4}%
\end{equation}

Notice that, when the discussion is directed from this point of view, the
derivation of the canonical noncommutativity deviates attention from a
fundamental point: the noncommutativity obtained as a consequence of
projection onto the lowest Landau level (which enforces the identification
between $\widehat{X}^{i}$ and $\hat{x}^{i}$) ought to be in reality assumed as
a property\ of the particle coordinate observables. This is the source of some
confusion in the literature on the Landau problem, and also in the
high-energy-physics articles on NCQFT that refer to it. For instance, take for
example the following assertion found in \cite{15.5} : ``An example of a
system where space-time coordinates do not commute is that of a particle in a
strong magnetic field''. One could ask\ how a strong magnetic field in a
laboratory could\ turn the spacetime noncommutative and thus destroy its
pointwise structure.

The canonical noncommutativity originated from string theory in \cite{3} is
based on an approximation which similar to the one of the lowest Landau level
just described.\ Consider open bosonic strings moving in a flat euclidean
space with metric $g_{\mu\nu},$ in the presence of a constant Neveu-Schwarz
$B$-field and with $Dp$ branes. The $B$-field probed by the open strings is
equivalent to a constant magnetic field on the branes, and it can be gauged
away in the directions transverse to the $Dp$ brane worldvolume. The
worldsheet action is
\begin{align}
S &  =\frac{1}{4\pi\alpha^{\prime}}\int_{\Sigma}\left(  g_{\mu\nu}%
\partial_{\eta}X^{\mu}\partial^{\eta}X^{\nu}-2\pi i\alpha^{\prime}B_{\mu\nu
}\epsilon^{\eta\lambda}\partial_{\eta}X^{\mu}\partial_{\lambda}X^{\nu}\right)
\label{5}\\
&  =\frac{1}{4\pi\alpha^{\prime}}\int_{\Sigma}g_{\mu\nu}\partial_{\eta}X^{\mu
}\partial^{\eta}X^{\nu}-\frac{i}{2}\int_{\partial\Sigma}B_{\mu\nu}X^{\mu
}\partial_{t}X^{\nu}\text{ \ ,}\nonumber
\end{align}
where $\alpha^{\prime}=l_{s}^{2}$, $\Sigma$ is the string worldsheet,
$\partial_{t}$ is a tangential derivative along the worldsheet boundary
$\partial\Sigma$ and the $X^{\mu}$ is the embedding function of the strings
into flat spacetime. If we consider the limit $g_{\mu\nu}\sim\left(
\alpha^{\prime}\right)  ^{2}\rightarrow0,$ keeping $B_{\mu\nu}$\ fixed
\cite{3}, the bulk kinetic terms of (\ref{5}) vanish. The worldsheet
theory\ in this limit is topological. All that remains are the boundary
degrees of freedom, which are governed by the action%

\begin{equation}
S=-\frac{i}{2}\int_{\partial\Sigma}B_{\mu\nu}X^{\mu}\partial_{t}X^{\nu
}.\label{6}%
\end{equation}
If one regards (\ref{6}) as a one dimensional action, and ignores the fact
that the $X^{\mu}\left(  t\right)  $ are the endpoints of a string, then
it\ can be considered as analogous to the action corresponding to the
Lagrangian of the Landau problem (\ref{3}). Under the approximation being
considered,\ the $X^{\mu}\left(  t\right)  $ may be regarded as operators
satisfying the canonical commutation relation
\begin{equation}
\lbrack\widehat{X}^{\mu},\widehat{X}^{\nu}]=\left(  \frac{i}{B}\right)
^{\mu\nu},\label{7}%
\end{equation}
which is identical to the one of (\ref{1}) by defining\ $\theta^{\mu\nu
}=\left(  1/B\right)  ^{\mu\nu}$.

The close similarity between this heuristic derivation\footnote{We abbreviated
the exposition to go directly to the point of interest of this work: the
analogy of the stringy noncommutativity to the one of the Landau problem. This
is an issue previously pointed out, for example, in\ \cite{1}, \cite{3}. A
detailed discussion on the derivation of (\ref{1}) would involve certain
subtleties, and may be found in those references.} and the one presented for
the Landau problem suggests that the noncommutativity achieved by the
approximation under consideration may be assumed to be of the particle
coordinate observables, instead of the spacetime coordinates. At this point,
it is interesting to quote that other noncommutative theories which emerged in
the stringy context, known as noncommutative dipole theories \cite{15.7}, were
presented that contain\ a noncommutativity also not of the spacetime
coordinates. In these theories noncommutativity is an inherent property\ of
the fields.

Although the great majority of papers on NCQFT consider the spacetime as
being\ ``pointless'', they use the basis of plane wave for their calculations
in momentum space (see for example \cite{1,1.5,3,7} and references therein)
and interpret the space of Weyl symbols \cite{13} as the physical position
space. This is equivalent to allow the localization of information at
individual points in the spacetime. The apparent contradiction in this
procedure is clearly shown to be absent if we interpret the stringy spacetime
that they are considering as commutative. It is important to keep in mind that
the distinction between the intrinsic spacetime noncommutativity and
the\ coordinate observable noncommutativity is not a question of metaphysics.
Since the calculation of the Green functions in the first case must employ the
procedure of averaging over localized states in the space of Weyl
symbols\ \cite{13} (which smears the pointwise information), the finiteness of
the theories, as well as the measurable physical quantities, will differ
according to the approach adopted for the noncommutativity.

Let us review the Weyl quantization procedure, considering stringy
noncommutativity from the new point of view.\ For the case under
consideration, it consists in the establishment of a map from the
noncommutative space of the particle coordinate observables to the space of
commutative coordinates by the use of the Moyal\ star product
\begin{align}
\left(  f\star g\right)  (x) &  =\frac{1}{(2\pi)^{n}}%
%TCIMACRO{\dint }%
%BeginExpansion
{\displaystyle\int}
%EndExpansion
d^{n}kd^{n}pe^{i(k_{\mu}+p_{\mu})x^{\mu}-\frac{i}{2}k_{\mu}p_{\nu}\theta
^{\mu\nu}}f(k)g(p)\nonumber\\
&  =\left.  e^{\frac{i}{2}\theta^{\mu\nu}\frac{\partial}{\partial\xi^{\mu}%
}\frac{\partial}{\partial\eta^{\nu}}}f(x+\xi)g(x+\eta)\right|  _{\xi=\eta
=0}.\label{8}%
\end{align}
According to this technique, the representation of a given NCQFT\ originally
valued on the space of noncommutative variables on the space of ordinary
commutative functions is achieved by the replacement of the noncommuting
variables in the arguments of the fields in the action by the commutative ones
and the ordinary product of the fields by the Moyal product. Since the
noncommutativity is assumed to be only of the particle coordinate observables,
what the Weyl procedure does is just to represent the field theory\ originally
valued on this noncommutative variables in function of the spacetime
coordinates and canonical momenta. This will become particularly clear later,
when we will consider NCQM.

Due to the property $\int d^{4}xA\star B=\int d^{4}x$ $AB$, the Hilbert (Fock)
space of a noncommutative field theory can be chosen to be the same as the one
of its commutative counterpart at the perturbative level \cite{26.5}, and the
noncommutativity is manifest only through the interaction terms of the action.
This gives us no surprise, since the difference of a commutative quantum field
theory and its noncommutative counterpart is only the presence of the\ strong
Neveu-Schwartz $B$-field in the background, acting to forbid the simultaneous
measurement of all the position coordinates of the particles. Of course, the
$B-$ field does this by interacting with the quantum fields, and thus its
presence must be encoded in the interaction terms of the action.

\section{NCQFT and the Hamilton-Jacobi Theory}

Here we discuss some aspects of the Schr\"{o}dinger representation for NCQFT
based on the canonical deformation of the algebra of the coordinate
observables. A study of NCQFT\ in the Heisenberg and path integral
formulations may be found in \cite{41}.

For simplicity, we will consider scalar field theories with action
\begin{equation}
S=\int d^{4}x\left\{  \frac{1}{2}\left(  \partial^{\mu}\varphi\partial_{\mu
}\varphi-\frac{m^{2}}{\hbar^{2}}\varphi^{2}\right)  -V_{\star}\left(
\varphi\right)  \right\}  \text{ \ ,}\label{10}%
\end{equation}
where $V_{\star}\left(  \varphi\right)  $ is the usual commutative interaction
part of the potential with the star product replacing the ordinary one. Since
the discussion\ here is general,\ and the procedure presented can be applied
to dimensions other than four, we will keep the potential as a generic real
analytic one. Of course, the question of renormalizability must be taken into
account when choosing the acceptable potentials \cite{30}. All the
quantization procedures of the usual formalism developed for the commutative
quantum field theory in the Schr\"{o}dinger representation \cite{31} are
similar in the noncommutative case under consideration, which presupposes that
$\theta^{0i}=0$. We briefly present the aspects of the procedure that will be
useful in this work.

The Hamiltonian corresponding to (\ref{10})\ is%

\begin{equation}
H=\int d^{3}x\left\{  \frac{1}{2}\left(  \pi^{2}+\left|  \nabla\varphi\right|
^{2}+\frac{m^{2}}{\hbar^{2}}\varphi^{2}\right)  +V_{\star}\left(
\varphi\right)  \right\}  ,\label{11}%
\end{equation}
where $\pi\left(  x\right)  =\dot{\varphi}\left(  x\right)  $ is the conjugate
field momentum. The operators $\widehat{\varphi}\left(  x\right)  $ and
$\widehat{\pi}\left(  x\right)  $ satisfy the canonical equal-time
commutators,
\begin{align}
\left[  \widehat{\varphi}\left(  \vec{x},t\right)  ,\widehat{\pi}\left(
\vec{y},t\right)  \right]   &  =i\hbar\delta\left(  \vec{x}-\vec{y}\right)
,\nonumber\\
\left[  \widehat{\varphi}\left(  \vec{x},t\right)  ,\widehat{\varphi}\left(
\vec{y},t\right)  \right]   &  =\left[  \widehat{\pi}\left(  \vec{x},t\right)
,\widehat{\pi}\left(  \vec{y},t\right)  \right]  =0\text{ .}\label{12}%
\end{align}

To work in a coordinate field representation,\ we shall\ consider a basis for
the Fock space where the operator $\widehat{\varphi}$ is time independent and
diagonal $\widehat{\varphi}\left(  \vec{x}\right)  \mid\phi\rangle=\phi\left(
\vec{x}\right)  \mid\phi\rangle$. In this basis, the state $\mid\Psi\rangle$
is represented by\ the time dependent wave functional $\Psi\left[
\phi,t\right]  =\langle\phi\mid\Psi\rangle$, and the momentum field operator
$\hat{\pi}\left(  \vec{x}\right)  $ by $-i\hbar\delta/\delta\phi\left(
\vec{x}\right)  $. The Schr\"{o}dinger equation is written as
\begin{equation}
i\hbar\frac{\partial}{\partial t}\Psi\left[  \phi,t\right]  =\int
d^{3}x\left\{  \frac{1}{2}\left(  -\hbar^{2}\frac{\delta^{2}}{\delta
\phi\left(  \vec{x}\right)  ^{2}}+\left|  \nabla\phi\right|  ^{2}+\frac{m^{2}%
}{\hbar^{2}}\phi^{2}\right)  +V_{\star}\left(  \phi\right)  \right\}
\Psi\left[  \phi,t\right]  .\label{18}%
\end{equation}

If we write the wavefunctional in its polar form $\Psi=R$ $\exp\left(
iS/\hbar\right)  $ and substitute it into (\ref{18}), we obtain
\begin{align}
\left[  i\hbar\frac{\partial R}{\partial t}-R\frac{\partial S}{\partial
t}\right]  e^{iS/\hbar} &  =\int d^{3}x\left\{  -\frac{\hbar^{2}}{2}\left[
\frac{\delta^{2}R}{\delta\phi^{2}}-\frac{R}{\hbar^{2}}\left(  \frac{\delta
S}{\delta\phi}\right)  ^{2}+i\left(  \frac{2}{\hbar}\frac{\delta R}{\delta
\phi}\frac{\delta S}{\delta\phi}+\frac{R}{\hbar}\frac{\delta^{2}S}{\delta
\phi^{2}}\right)  \right]  \right\}  e^{\frac{i}{\hbar}S}\nonumber\\
&  +\int d^{3}x\left[  \frac{1}{2}\left|  \nabla\phi\right|  ^{2}+\frac{m^{2}%
}{2\hbar^{2}}\phi^{2}+V_{\star}\left(  \phi\right)  \right]  Re^{iS/\hbar
}\label{18.5}%
\end{align}
Dividing (\ref{18.5})\ by $\Psi$ and\ separating the real part, we find
\begin{equation}
\frac{\partial S}{\partial t}+\int d^{3}x\left\{  \frac{1}{2}\left[  \left(
\frac{\delta S}{\delta\phi}\right)  ^{2}+\left|  \nabla\phi\right|  ^{2}%
+\frac{m^{2}}{\hbar^{2}}\phi^{2}\right]  +V\left(  \phi\right)  \right\}
+V_{nc}[\phi]+Q_{K}[\phi]=0,\label{19}%
\end{equation}
where
\begin{equation}
V_{nc}[\phi]=\int d^{3}x\left\{  \operatorname{Re}\left[  V_{\star}\left(
\phi\right)  \right]  -V\left(  \phi\right)  \right\}  ,\label{19.2}%
\end{equation}
and
\begin{equation}
Q_{K}[\phi]=-\frac{\hbar^{2}}{2R}\int d^{3}x\frac{\delta^{2}R}{\delta
\phi\left(  \vec{x}\right)  ^{2}}.\label{19.4}%
\end{equation}
Equation (\ref{19}) is the ordinary Hamilton-Jacobi equation for the scalar
field with two additional potentials $V_{nc}$ and $Q_{K}.$

The complete information contained in the functional Schr\"{o}dinger equation
is extracted only if we consider (\ref{19}) and the other coupled equation
that is obtained from the decomposition of (\ref{18}) into its real and
imaginary parts. Multiplying (\ref{18.5})\ by\ $2R/\hbar\exp\left(
-iS/\hbar\right)  $,\ taking the imaginary part, and doing\ some
simplification using the properties of the star product, the equation obtained
reads
\begin{equation}
\frac{\partial R^{2}}{\partial t}+\int d^{3}x\frac{\delta}{\delta\phi}\left(
R^{2}\frac{\delta S}{\delta\phi}\right)  =0.\label{21}%
\end{equation}
This expression must be interpreted as a continuity equation for the
probability density $R^{2}[\phi(\vec{x}),t]$ that the field configuration be
$\phi(\vec{x})$ at time $t$. Notice that (\ref{19}) and (\ref{21}) constitute
a set of nonlinear\ coupled equations. In practice, it is convenient to solve
directly (\ref{18})\ and then obtain $R$ and $S$ from the wavefunctional
$\Psi$.\ The advantage of decomposing the functional Schr\"{o}dinger equation
in a more complex system of two coupled equations is for the sake of its
physical interpretation.

In what follows, we shall use the generalized Hamilton-Jacobi formalism
introduced to perform an analysis of\ the conditions for a system achieve its
classical and commutative limits. Before going on, however, it is necessary to
define precisely what we mean for achieving these limits.

Finding the classical or commutative limit of a system is to establish the
conditions that must be imposed on the environment, quantum numbers or
physical constants in order to enforce the system to assume the behavior
identical to the one of a classical or commutative analog. These conditions
can be determined by analyzing equations (\ref{19}) and (\ref{21}). When we
consider a solution to the equations (\ref{19}) and (\ref{21}), and imagine it
as being substituted on them, the original set of equations must now be
understood as a set of identities that are being trivially satisfied. The
classical or commutative limits are achieved when the conditions are such that
this set of equations (seen as a set of identities)\ assume the classical form
or the quantum commutative form. Notice that, according to this criterion, the
establishment of the adequate prescription for the achievement of the limits
of each individual system must be done by accounting for the properties that
characterize its physical state. The procedure proposed will thus lead to the
determination of state dependent criteria, thought they can be written in an
universal and compact form, as will be shown later.

The first interesting fact to be noticed once one adopts the prescription
suggested here, is that the achievement of the classical limit will, in
certain cases, occur under conditions that differ form the naive one,
$\hbar\rightarrow0,$ usually employed in many textbooks. Here, and in what
follows,\ the symbol $``A\rightarrow B"$ must be understood as taking a limit
such that $A-B$ is sufficiently small to be neglected in comparison with the
other\ quantities which are relevant to the physical system under
consideration. For example, $\hbar\rightarrow0$ may be assumed to\ mean that
$\hbar\ll S_{0}$, where $S_{0}$ is some characteristic action of the
corresponding classical motion of the system \cite{22.5,31.5}. In\ the
achievement of the classical limit starting from (\ref{19}) and (\ref{21}), we
must\ take into account the fact that $R$ (and$\ S$) in principle depends on
$\hbar$, and hence in general the condition $Q_{K}\rightarrow0$ is not
satisfied always when $\hbar\rightarrow0$. A detailed discussion about the
fallacy in adopting $\hbar\rightarrow0$ as the universal criterion for the
classical limit\ in the context of ordinary quantum mechanics and field theory
containing examples\ is found in \cite{18,32} and references therein.

An analogous reasoning can be applied when we consider the commutative limit.
Although $V_{nc}\rightarrow0$ when $\theta^{ij}\rightarrow0$, as can be seen
from (\ref{8}) and (\ref{19.2}), this criterion may not be valid for the
achievement of the commutative limit. This is a consequence of the fact
that$\ Q_{K}$ may contain contributions generated by the noncommutativity that
do not depend on $\theta^{ij}$. On that case, taking $\theta^{ij}\rightarrow0$
will not vanish them, do not conducting to the commutative limit. Moreover,
since in principle the dependence of $Q_{K}$ on $\theta^{ij}$ is arbitrary and
totally state dependent, the possibility of $Q_{K}$ blowing-up when
$\theta^{ij}\rightarrow0$ must not be discarded. In reality this is expected
to occur in some models.

When we consider a differential equation which is deformed depending on a set
of parameters (for example $\theta^{ij}$), the set of admissible solutions is
richer than the original one corresponding to the undeformed equation. If we
smoothly undo the deformation,\ part of the expanded set of solutions will of
course suffer a process of homotopy, and in the end of it will be contained in
the set of solutions of the original undeformed equation. But part of the same
set will not present the same behavior, and may become unaltered, go to
another limit outside the set of undeformed solutions, or blow-up.

In fact, it has been verified that in many of the perturbative loopwise
calculations of NCQFT, some peculiar phenomena occurs when $\theta
^{ij}\rightarrow0$, like UV/IR mixing \cite{7}, which is responsible for the
non-analytic behavior in the $\theta^{ij}\rightarrow0$ limit, the blowing-up
of the self-energy in some models \cite{32.7}, non-reduction of some gauge
theories to the commutative counterparts\ \cite{32.8}, among others
\cite{1.5}. In the quantum-mechanical context, it is interesting to quote the
blowing-up of the energy spectra \cite{32.9} and the appearance of
wavefunctions representing extremely localized states \cite{8,32.95}. This
states are particularly interesting as they present nonperturbative effects in
$\theta^{ij}$ at small distances. In addition,\ if one starts considering
these (normalizable) wavefunctions\ and consider the limit where $\theta^{ij}$
vanishes, they become Dirac delta functions, which are non-normalizable
\cite{32.95}.

In what follows, we will turn operational the technique proposed for the
determination of the classical and commutative limits by showing how it can be
applied to perform an analysis of NCQFT. The origin of the quantum
contributions on (\ref{19}) is traced back to\ the action of the field
momentum in its differential representation, $\hat{\pi}\left(  \vec{x}\right)
=-i\hbar\delta/\delta\phi\left(  \vec{x}\right)  ,$ on the wavefunctional
$\Psi=R\exp\left(  iS/\hbar\right)  $ in\ (\ref{18}). When this equation is
divided by $\Psi$ and the real part is taken,\ the quantum Hamilton-Jacobi
equation (\ref{19}) is originated containing terms that cannot appear on the
classical Hamilton-Jacobi equation, where the field momentum is represented by
$\pi\left(  \vec{x}\right)  =\delta S/\delta\phi\left(  \vec{x}\right)  $.
These terms are nonclassical, and thus must be responsible by the quantum
effects. This is a\ simple and safe criterion for the identification of the
quantum contributions.

In practice, it is not necessary to follow all the steps and write the
Hamilton-Jacobi equation containing all its terms.\ When one wants to identify
and compute\ the quantum effects corresponding to an specific part of the
Schr\"{o}dinger Hamiltonian, the procedure\ can be\ applied\ directly to the
term under consideration to extract its quantum contributions, which are
grouped to constitute its associated quantum potential. In equation
(\ref{18}), for example, the only term that has derivatives (here functional)
acting on $\Psi$ is kinetic one. Thus, there will be only one quantum
potential, which must be associated with this term and is given by
\begin{align}
Q_{K} &  =\operatorname{Re}\left(  -\frac{\hbar^{2}}{2}\frac{1}{\Psi}\int
d^{3}x\frac{\delta^{2}\Psi}{\delta\phi^{2}}\right)  -\left(  \frac{\hbar^{2}%
}{2}\int d^{3}x\left(  \frac{\delta S}{\delta\phi}\right)  ^{2}\right)
\nonumber\\
&  =-\frac{\hbar^{2}}{2R}\int d^{3}x\frac{\delta^{2}R}{\delta\phi\left(
\vec{x}\right)  ^{2}}.\label{22}%
\end{align}
This reproduces (\ref{19.4}), and is the expression used to define
the\ quantum potential in the de Broglie-Bohm interpretation of\ quantum
theory\footnote{Usually, the polar decomposition of the wavefunctional and the
associated Hamilton-Jacobi formalism are explored by the Bohmian community in
the direction of establishing an ontological meaning for quantum field theory,
with an additional assumption about the objective existence of quantum fields
independent of the act of observation \cite{18,32}. This assumption, however,
is not being\ considered in this work. The most general case where we intend
that this method could be applied may not yet admit an ontological
interpretation.} \cite{32}.

It is possible to associate a quantum potential to each term or, according to
the necessity, a group of terms involving derivatives in the Hamiltonian. To
suppress its corresponding quantum effects, it is sufficient to impose the
cancelation of its corresponding quantum potential as an equation. The
classical limit of the theory is found when the sum of\ all the quantum
potentials vanish. In this case the Hamilton-Jacobi equation looses its
dependence on $R$ and\ on second and higher order derivatives of $S$,
decouples from (\ref{21}) and\ assumes its classical form, which is an
equation exclusively for $S$. This is a general result, totally independent of
the specific form of equation\ (\ref{19}). By definition, the classical
contributions are the ones obtained by the replacement $\pi\left(  \vec
{x}\right)  =\delta S/\delta\phi\left(  \vec{x}\right)  $, and thus do not
contain any derivative of $R$ that could spoil the decoupling process of the
Hamilton-Jacobi equation. Once (\ref{19}) and (\ref{21}) assume a form
characteristic of the classical physics, the physical content of the system is
the same of the classical field theory, and therefore the observable
quantities, as well as the equations of evolution of their averages, will be
indistinguishable of the ones of classical physics.

Having determined the criterion for the classical limit as $Q_{K}\rightarrow
0$, let us concentrate our attention on the criterion for the commutative one.
Since equation (\ref{21}) is still in the ordinary commutative classical form,
the discussion about how to achieve the commutative limit reduces to the
analysis of the potentials $Q_{K}$ and $V_{nc}$ of (\ref{19}). One safe way
for the identification of the noncommutative effects present in $Q_{K}$ is to
calculate the same quantum potential with respect to the analog wavefunctional
corresponding to the associated commutative field theory and compare with it.
This is done by defining the commutative quantum potential as
\begin{equation}
Q_{c}=-\frac{\hbar^{2}}{2R}\int d^{3}x\frac{\delta^{2}R_{c}}{\delta\phi\left(
\vec{x}\right)  ^{2}},\label{23}%
\end{equation}
where $R_{c}=\sqrt{\Psi_{c}^{\ast}\Psi_{c}}$ and $\Psi_{c}$ is the solution of
(\ref{18}) with $\theta^{ij}=0$, that is , the solution of the commutative
counterpart. The noncommutative contributions contained in $Q_{K} $ can thus
be attributed to the functional
\begin{equation}
Q_{nc}=Q_{K}-Q_{c},\label{24}%
\end{equation}
which we define as the noncommutative quantum potential.

When $V_{nc}+Q_{nc}\rightarrow0$,\ (\ref{19}) becomes similar to the
Hamilton-Jacobi equation corresponding to the conventional quantum field
theory, assuring the noncommutative/commutative passage. This can happen in
two ways:

1) $V_{nc}+Q_{nc}\rightarrow0$\ is achieved by taking the limit $\theta
^{ij}\rightarrow0$ directly without any obstruction caused by the quantum
effects represented by $Q_{nc}$. In this case the $\theta^{ij}$ disappear
completely from all terms of the Hamilton-Jacobi\ equation. This is the most
common approach and easy to understand in comparison with classical
noncommutative field theory, where $Q_{nc}$\ is absent the commutative limit
is usually achieved\ by taking $\theta^{ij}\rightarrow0$.

2) $V_{nc}+Q_{nc}\rightarrow0$ is achieved by varying a physical constant or
parameter of the system, keeping $\theta^{ij}$ unaltered. This approach is
less common, but may be useful specially in cases where the first is not
possible to be realized.

Two examples of application of the Hamilton-Jacobi formalism in the quantum
mechanical context are presented an worked out in the end of the article.

\section{ Noncommutative Quantum Mechanics}

Here, we discuss the implications of considering the coordinate observables of
the particles as operators satisfying (\ref{1}) for quantum mechanics. We
shall show how to work in the Schr\"{o}dinger formulation\ with a consistent
interpretation for the wavefunction and how the Hamilton-Jacobi formalism of
the previous section can be applied to give information about the formal
structure of NCQM in comparison with the one of NCQFT.\ 

\subsection{Establishing the Foundations}

As in the commutative theories, NCQM can be derived from NCQFT in its\ the low
energy limit. A rigorous treatment for this derivation was done P. M. Ho and
H. C. Kao in \cite{33}. For chargeless particles their noncommutative
Schr\"{o}dinger\ equation reduces to
\begin{equation}
i\hbar\frac{\partial\Psi(x^{i},t)}{\partial t}=-\frac{\hbar^{2}}{2m}\nabla
^{2}\Psi(x^{i},t)+V(x^{i})\star\Psi(x^{i},t).\label{26}%
\end{equation}
By using the properties of the star product, the potential term of the
Schr\"{o}dinger equation can also be written as
\begin{equation}
V(x^{i})\star\Psi(x^{i},t)=V\left(  x^{i}+\frac{i}{2}\theta^{ij}\partial
_{j}\right)  \Psi(x^{i},t).\label{27}%
\end{equation}
The Hamiltonian is then%

\begin{equation}
H=\frac{-\hbar^{2}}{2m}\nabla^{2}+V\left(  x^{i}+\frac{i}{2}\theta
^{ij}\partial_{j}\right)  .\label{28}%
\end{equation}
Thus, the noncommutative quantum mechanics is physically equivalent to a
commutative quantum mechanics with new momentum dependent interactions,
dictated by the replacement $x^{i}\rightarrow x^{i}-\theta^{ij}\hat{p}%
_{j}/2\hbar$ in the potential term. This can be assumed to be the Weyl
correspondence for quantum mechanics, and has been widely used in the
literature (see, for example \cite{27}). However, the complete investigation
of all of its consequences has not ever been done.

To understand physics that lies behind the correspondence above, we again
appeal to the Landau problem, this time do not considering the projection onto
the lowest Landau level. An analogy between the physical variables suggests
that the shift $x^{i}\rightarrow x^{i}-\theta^{ij}\hat{p}_{j}/2\hbar$ caused
by the external Neveu-Schwartz $B$ field is similar to the shift $\hat{p}%
^{i}\rightarrow\hat{p}^{i}-e\hat{A}^{i}$, which is\ caused by an external
magnetic field in ordinary quantum mechanics. Since in the Landau problem the
physical momentum is $\hat{\pi}^{i}=\hat{p}^{i}-e\hat{A}^{i},$ according to
the analogy, the physical position must be\footnote{The association of the
$\widehat{X}^{i}$ with the observables corresponding to the\ physical
coordinates of the particles was also\ proposed in \cite{28}, but following
the same line of \cite{27}, which interpret the canonical positions $x^{i}$
just as auxiliary variables, do not having any physical interpretation.}
$\widehat{X}^{i}=x^{i}-\theta^{ij}\hat{p}_{j}/2\hbar$. This intuitively
explains why the position observables must remain noncommuting after the Weyl
correspondence is applied. This map is nothing but a representation of the
noncommuting position observables in terms of the canonical momentum $\hat
{p}^{i}$ and canonical coordinates $x^{i}$, the latter having the meaning of
being points of the real physical space.

Some physical intuition on the meaning of the canonical position variables can
be obtained by considering the dipole picture \cite{35}. For the case of the
NCQM under consideration, this consists in considering that, instead of
a\ particle, the elementary object of the theory is a half dipole\footnote{In
the context of this work, which considers point particles as the true
elementary constituents of matter, the dipoles are assumed to be fictitious.}
whose extent is proportional to its momentum, $\Delta x^{i}=\theta^{ij}%
p_{j}/2\hbar$. One of the endpoints of the\ dipole carries its mass and is
responsible for its interactions. The other extreme is empty. According to
this intuitive view, the change of variables $X^{i}=x^{i}-\theta^{ij}%
p_{j}/2\hbar$ corresponds to a change of coordinates of the interacting
extreme of the dipole $X^{i}$,\ where the corresponding physical particle is
located, to its\ empty one $x^{i}$. Therefore, the effect of the background
field, which acts to\ forbid the coordinates to commute, is compensated in
this new commutative\ coordinate system obtained by change of variable that is
dependent on the momentum and background intensity (remember $\theta
^{ij}=(1/B)^{ij}$). Although\ the simultaneous measurement of the observables
$\hat{x}^{i}$ is in principle possible, it is not sufficient for the
determination of the physical position of the associated particle. As $\hat
{x}^{i}$ and $\hat{p}^{i}$ do not commute, the knowledge of the three $x^{i}$
forbids the simultaneous determination of the three $p^{i}$ and thus the
physical location of the particle.

The Hilbert space of states of noncommutative quantum mechanics is assumed to
be the same of the commutative one \cite{27}, which is a remnant from a result
of field theory discussed in the Second Section. By knowing the Hilbert space
of states and the Schr\"{o}dinger evolution equation for the wavefunction, it
remains to show that this wavefunction admits the definition of a density of
probability. The usual definition of probability density%

\begin{equation}
\rho=\Psi^{\ast}\Psi=\left|  \Psi\right|  ^{2}\label{29}%
\end{equation}
can be employed with some care in interpreting its meaning. The wavefunction
here is valued on the canonical coordinates $x^{i}$, rather than on
eigenvalues of the physical position observables $\widehat{X}^{i}$. Thus
$\rho(\vec{x},t)d^{3}x$ must be interpreted as the probability of finding the
canonical coordinate of the particle in the volume $d^{3}x$ around the point
$\vec{x}$ at time $t$.

To extract information on\ the physical coordinate position of the particles
from the wavefunction, it is necessary to expand it in eigenfunctions of the
position operators $\widehat{X}^{i}=x^{i}+i\theta^{ij}\partial_{j}/2$. Of
course, since the $\widehat{X}^{i}$ do not commute, it is impossible to find a
basis for all the spatial directions simultaneously. This is the simple
manifestation of the fact that, although the spacetime is endowed with a
pointwise structure, it is impossible to localize a particle on a given point
of it in a measurement. The background Neveu-Schwartz field always acts
conspiring against any attempt in this direction, enforcing the results of the
measurements to obey the uncertainty relation
\begin{equation}
\Delta X^{i}\Delta X^{j}\geq\left|  \theta^{ij}\right|  /2.\label{29.5}%
\end{equation}
This solves the open question about the possibility of the existence of the
wavefunction and how to interpret it \cite{27,27.5} in the context of the
string- inspired noncommutativity. Had we considered the wrong interpretation
of the canonical position observables $\hat{x}^{i}$ as the physical ones, the
uncertainty relation (\ref{29.5}) would be lost in the Schr\"{o}dinger
formulation. Of course, due to the smallness expected for $\theta^{ij},$ in
practice\ the noncommutative effects are not possible to be perceived by a
direct verification of (\ref{29.5}), which would demand an energy to work on a
length scale far beyond the limits of validity of nonrelativistic quantum
mechanics, but through its indirect consequences for other observable
quantities, like atomic energy spectrum \cite{27,27.3}, phase shifts on the
Aharonov -Bohm effect \cite{39}, etc.

By differentiating (\ref{29}) and using (\ref{26}) we obtain
\begin{equation}
\frac{\partial\rho}{\partial t}=\frac{i\hbar}{2m}\nabla\cdot\left(  \Psi
^{\ast}\nabla\Psi-\Psi\nabla\Psi^{\ast}\right)  -\frac{i}{\hbar}\left(
\Psi^{\ast}V\star\Psi-\Psi\left(  V\star\Psi\right)  ^{\ast}\right)
,\label{30}%
\end{equation}
where the term$\ i\hbar\left(  \Psi\nabla\Psi^{\ast}-\Psi^{\ast}\nabla
\Psi\right)  /2m$ is well known from ordinary quantum mechanics and
corresponds in that case to the current density $\vec{J}$. Notice that, since
the canonical coordinates $x^{i}$ do not represent the physical coordinates of
the particles, the fact that the usual continuity equation$\ \partial
\rho/\partial t+\nabla\cdot\vec{J}=0$ is not satisfied\ does not mean that
there is no local conservation law of probability. Such form of continuity
equation would be expected only if $\left|  \Psi(x^{i},t)\right|  ^{2}d^{3}x$
had the meaning of probability of finding the particle in a volume $d^{3}x$
around the point $\vec{x}$ at time $t$.

Actually, the extra term that appeared in (\ref{30}) due to the
noncommutativity is not responsible for any inconsistency of the theory, but
its role must be understood. One way to justify its presence is to notice that
it is essential in order to assure that the \textit{equivariance} property
\cite{22.5} is satisfied by $\rho=\left|  \Psi\right|  ^{2}$. This means that
if $\rho(x^{i},t_{0})=\left|  \Psi(x^{i},t_{0})\right|  ^{2}$ at some time
$t_{0}$, then $\rho(x^{i},t)=\left|  \Psi(x^{i},t)\right|  ^{2}$ for all $t$.
In other words, $\rho(x^{i},t)$ preserves its form as a functional of
$\Psi(x^{i},t)$ for all times. This property is trivially satisfied because
the starting point for the derivation of the equation (\ref{30}) was exactly
the definition of $\rho(x^{i},t)=\left|  \Psi(x^{i},t)\right|  ^{2}$ as the
density of probability for an arbitrary time and the noncommutative
Schr\"{o}dinger equation. Thus it is the noncommutative Schr\"{o}dinger
equation itself that\ enforces the appearance of the noncommutative term in
the local probability\ conservation law for a question of consistency. This
term could called ``\textit{equivariance} correction'', and be denoted by
$\Sigma_{\theta}$. It should be expected, however, that when integrated over
all the space $\Sigma_{\theta}$ vanish, as occurs with the term containing the
divergence. This is a\ necessary condition\ for the global conservation of
probability. By integrating equation (\ref{30}) over the space\ and using the
properties of the star product\ it is easy to verify that
\begin{equation}
\int\left(  \frac{\partial\rho}{\partial t}+\nabla\cdot\left[  \frac{i\hbar
}{2m}\left(  \Psi\nabla\Psi^{\ast}-\Psi^{\ast}\nabla\Psi\right)  \right]
\right)  d^{3}x=-\frac{i}{\hbar}\int\left[  \Psi^{\ast}V\star\Psi-\Psi\left(
V\star\Psi\right)  ^{\ast}\right]  d^{3}x=0,\label{32}%
\end{equation}
thanks to the antisymmetry of $\theta^{\mu\nu}$.

\subsection{Analyzing From Hamilton-Jacobi Point of View}

The Hamilton-Jacobi formalism associated with the NCQM is found by applying
the same procedure previously discussed when considering the NCQFT. We write
the wavefunction in its polar form $\Psi=R\,e^{iS/\hbar}$, substitute into
equation (\ref{26}), and separate its real and imaginary parts. For the real
part we obtain%

\begin{equation}
\frac{\partial S}{\partial t}+\frac{\left(  \nabla S\right)  ^{2}}%
{2m}+V+V_{nc}+Q_{K}+Q_{I}=0.\label{39}%
\end{equation}
The three new potential terms are defined as
\begin{equation}
V_{nc}=V\left(  x^{i}-\frac{\theta^{ij}\partial_{j}S}{2\hbar}\right)
-V\left(  x^{i}\right)  ,\label{40}%
\end{equation}%
\begin{equation}
Q_{K}=\operatorname{Re}\left(  -\frac{\hbar^{2}}{2m}\frac{\nabla^{2}\Psi}%
{\Psi}\right)  -\left(  \frac{\hbar^{2}}{2m}\left(  \nabla S\right)
^{2}\right)  =-\frac{\hbar^{2}}{2m}\frac{\nabla^{2}R}{R},\label{41}%
\end{equation}
and
\begin{equation}
Q_{I}=\operatorname{Re}\left(  \frac{V\left(  x^{i}+\frac{i}{2}\theta
^{ij}\partial_{j}\right)  \Psi}{\Psi}\right)  -V\left(  x^{i}-\frac
{\theta^{ij}}{2\hbar}\partial_{j}S\right)  .\label{42}%
\end{equation}
$V_{nc}$ is the\ potential that accounts for the noncommutative
classical\ interactions, while $Q_{K}$ and $Q_{I}$ for the quantum effects.
The origin of the $Q_{K}$ potential comes from\ the kinetic term of the
Hamiltonian,\ closely related to the case of NCQFT discussed before. Its
definition is identical to the one of the quantum potential in ordinary
Bohmian mechanics \cite{18}, and here we will call it the kinetic quantum
potential. The remaining term, $Q_{I}$, is the potential that accounts for the
quantum effects that come from the interaction term, and thus should be called
interaction quantum potential.

The imaginary part of the Schr\"{o}dinger equation yields
\begin{equation}
\frac{\partial R^{2}}{\partial t}+\nabla\cdot\left(  R^{2}\frac{\nabla S}%
{m}\right)  +\frac{2R}{\hbar}\operatorname{Im}\left(  e^{-\frac{i}{\hbar}%
S}V\star(Re^{\frac{i}{\hbar}S})\right)  =0,\label{44}%
\end{equation}
which\ is identical to equation (\ref{30}) for probability conservation, now
written in function of the $R$- and $S$-fields.

Returning to the equation (\ref{39}), we see that while$\ V_{nc}$\ and $Q_{K}
$\ have their partners in NCQFT, $Q_{I}$, on the other hand, has no analog
when a comparison is done with field theory. It could be written as
$Q_{I}(x^{i},S,\partial^{i}S,\partial^{ij}S,\partial^{ijk}S,...,R,\partial
^{i}R,\partial^{ij}R,\partial^{ijk}R,...)$, keeping in mind, of course, that
the higher derivative interactions come from a star product of $V$ with the
$\Psi$ field, and therefore $Q_{I}$ is not exactly an arbitrary function of
$x^{i},S,R$ and their partial derivatives. When $V$ is polynomial, $Q_{I}$
contains a finite number of derivatives, otherwise it will be infinite series
containing derivatives of all orders, which characterizes a interaction
nonlocal in space. This case can still be analyzed, at least perturbatively. A
general\ discussion of how to handle mathematically nonlocal particle
mechanics and field theory may be found in \cite{36,36.5}, and in the context
of noncommutative theories in \cite{37}. The reason for the absence of the
interaction\ quantum potential $Q_{I}$ in the case of NCQFT is that in that
case there\ is no star product between the potential and the wavefunctional in
the Schr\"{o}dinger equation. In the case of\ quantum field theory, the
potential term is valued on the star algebra and its product with the
wavefunctional is the ordinary one, while, in the quantum mechanical case, the
potential is valued on the ordinary algebra and multiplies the wave function
with the star product.

Notice that, in case of considering quantum field theory with an external
potential, it will be incorporated in the quadratic term of the Hamiltonian
with the star product, but\ its product with the wavefunctional will still\ be
the ordinary one.\ This point sets a difference between the implications of
noncommutativity for high and low energy phenomena. At least in what concerns
the structure of the equations in the Hamilton-Jacobi formalism, the
complexity and richness of the new physics originated from the
noncommutativity may be manifest in a more impressive\ way\ in the low energy processes.

\subsection{The Classical and Commutative Limits}

The most quoted criterion for finding the classical limit of a
quantum-mechanical systems is $\hbar\rightarrow0$. As pointed out when
discussing the NCQFT, the adoption of this criterion\ as universal for the
classical limit\ has\ several problems. For the determination of the
commutative limit, the most employed criterion is $\theta^{ij}\rightarrow0$
\cite{1,1.5}. However, arguments for its failure\ in the most general case
were presented in the framework of NCQFT in the Third Section. The arguments
presented there are still valid and reinforced here, where they become yet
stronger by the presence of the additional\ quantum potential $Q_{I}$ and the
\textit{equivariance} correction $\Sigma_{\theta}$. Among other criteria
commonly used for the classical limit are high quantum numbers ($n\rightarrow
\infty$), large mass ($m\rightarrow\infty$), and short de Broglie wavelength.
All of these criteria are not universal, and can be obtained from the
application for particular systems of the most general and systematic method
based on the generalized\ Hamilton-Jacobi formalism. The latter,\ according to
the case under consideration, provides the best choice as the parameter to vary.

The conditions for the attainment of the classical and commutative limits of a
quantum-mechanical system can be obtained easily from the examination of the
system of equations constituted by (\ref{39}) and (\ref{44}). For
simplicity,\ we shall restrict our considerations to the case where
$\Sigma_{\theta}\rightarrow0$, which will be the one of interest for the
examples that follow.\ According to the prescription for the identification of
the quantum effects proposed in Section 3, and applied in this subsection to
NCQM, all the quantum manifestations of a given noncommutative
quantum-mechanical system must be governed by $Q_{K}$ and $Q_{I}$ Thus, in the
limit where $Q_{K}+Q_{I}\rightarrow0$ the system is expected to present a
classical behavior.

By looking for the criteria for the commutative limit, however, we see that
the conditions that must be imposed on the set $\{V_{nc},Q_{K},Q_{I}\}$\ are
less trivial. While useful for the formal analysis comparing NCQM with NCQFT,
the\ grouping of the quantum contributions in the two potentials $Q_{K}$ and
$Q_{I}$ is not convenient for the identification of the noncommutative quantum
effects. This can be done by following an approach similar to the one adopted
for NCQFT. We define the noncommutative quantum contributions by
\begin{equation}
Q_{nc}=Q_{K}+Q_{I}-Q_{c},\label{44.2}%
\end{equation}
where
\[
Q_{c}=-\frac{\hbar^{2}}{2m}\frac{\nabla^{2}R_{c}}{R_{c}}\text{ \ ,\ \ \ }%
R_{c}=\sqrt{\Psi_{c}^{\ast}\Psi_{c}}.
\]
$\Psi_{c}$ is the wavefunction obtained from the commutative Schr\"{o}dinger
equation containing the usual potential $V(x^{i})$, that is, the equation
obtained by making $\theta^{ij}=0$ on (\ref{26}) before solving it. The
conditions to assure that (\ref{39}) and (\ref{44}) assume a form identical to
the one corresponding to a commutative system are
\begin{equation}
V_{nc}+Q_{nc}\rightarrow0.\label{44.4}%
\end{equation}
The condition for the classical limit can now be written as $Q_{c}%
+Q_{nc}\rightarrow0,$ and (\ref{39}) as
\begin{equation}
\frac{\partial S}{\partial t}+\frac{\left(  \nabla S\right)  ^{2}}%
{2m}+V+V_{nc}+Q_{c}+Q_{nc}=0.\label{44.6}%
\end{equation}

\section{Simple Applications}

\subsection{Noncommutative Harmonic Oscillator}

Here we show a simple application of the ideas presented in the latter
section. Consider a\ two dimensional noncommutative harmonic oscillator. In
two dimensions, (\ref{1}) can be written as
\begin{equation}
\lbrack\widehat{X}^{\mu},\widehat{X}^{\nu}]=i\theta\epsilon^{\mu\nu
}.\label{44.8}%
\end{equation}
The Hamiltonian of the system is given by
\begin{equation}
H=\frac{1}{2m}\left(  p_{x}^{2}+p_{y}^{2}\right)  +\frac{1}{2}mw^{2}\left[
\left(  x-\frac{\theta}{2\hbar}p_{y}\right)  ^{2}+\left(  y+\frac{\theta
}{2\hbar}p_{x}\right)  ^{2}\right]  ,\label{45}%
\end{equation}
where $m$ is the particle mass and $w$ the frequency of the associated commutative\ oscillator.

The corresponding Schr\"{o}dinger equation in polar coordinates is
\begin{align}
i\hbar\frac{\partial\Psi_{\theta}\left(  r,\varphi,t\right)  }{\partial t} &
=H_{\theta}\Psi_{\theta}\left(  r,\varphi,t\right) \nonumber\\
&  =-\frac{\hbar^{2}}{2m}\left(  1+\left(  \frac{mw\theta}{2\hbar}\right)
^{2}\right)  \left(  \partial_{r}^{2}+\frac{1}{r}\partial_{r}+\frac{1}{r^{2}%
}\partial_{\varphi}^{2}\right)  \Psi_{\theta}\left(  r,\varphi,t\right)
\nonumber\\
&  +\left(  i\frac{m}{2}\theta w^{2}\partial_{\varphi}+\frac{m}{2}w^{2}%
r^{2}\right)  \Psi_{\theta}\left(  r,\varphi,t\right)  .\label{46}%
\end{align}
By plugging $\Psi_{\theta}\left(  r,\varphi,t\right)  =e^{-iEt/\hbar}%
\psi_{\theta}\left(  r,\varphi\right)  $ into (\ref{46}) we obtain the
eigenvalue equation $H_{\theta}\psi_{\theta}\left(  r,\varphi\right)
=E_{\theta}\psi_{\theta}\left(  r,\varphi\right)  $. The solution of this
eigenvalue equation\ is straightforward (see, for example \cite{38}), and
gives
\begin{equation}
\psi_{\theta}\left(  r,\varphi\right)  =\psi_{n,\alpha,\theta}\left(
r,\varphi\right)  =\left(  -1\right)  ^{n}\sqrt{\frac{n!\tilde{\zeta}}%
{\pi\left(  n+\left|  \alpha\right|  \right)  !}}\exp\left(  -\frac
{\tilde{\zeta}r^{2}}{2}\right)  \left(  \sqrt{\tilde{\zeta}}r\right)
^{\left|  \alpha\right|  }L_{n,\theta}^{^{\left|  \alpha\right|  }}\left(
\tilde{\zeta}r^{2}\right)  e^{i\alpha\varphi},\label{47}%
\end{equation}
where $L_{n,\theta}^{^{\left|  \alpha\right|  }}\left(  \tilde{\zeta}%
r^{2}\right)  $ are the Laguerre polynomials
\begin{equation}
L_{n,\theta}^{^{\left|  \alpha\right|  }}\left(  \tilde{\zeta}r^{2}\right)
=\overset{n}{\underset{l=0}{\sum}}\left(  -1\right)  ^{n}\left(
\begin{array}
[c]{c}%
n+\left|  \alpha\right| \\
n-l
\end{array}
\right)  \frac{\left(  \tilde{\zeta}r^{2}\right)  ^{l}}{l!},\text{
\ \ \ \ \ }\tilde{\zeta}^{2}=\frac{\left(  \frac{mw}{\hbar}\right)  ^{2}%
}{1+\left(  \frac{mw\theta}{2\hbar}\right)  ^{2}}\text{\ },\label{48}%
\end{equation}
$n=0,1,2,$...is the principal quantum number, and $\alpha=0,\pm1,\pm2$... is
the angular canonical momentum quantum number. The corresponding energy
spectrum is
\begin{equation}
E_{n,\alpha,\theta}=2\hbar w\left(  1-\left(  \frac{mw\theta}{2\hbar}\right)
^{2}\right)  ^{1/2}\left(  n+\frac{\left|  \alpha\right|  +1}{2}\right)
-\frac{m\theta w^{2}\alpha}{2}.\label{49}%
\end{equation}

Let us consider the state where $n=0$, whose wavefunction is
\begin{equation}
\Psi_{\theta}\left(  r,\varphi,t\right)  =\sqrt{\frac{\tilde{\zeta}}%
{\pi\left|  \alpha\right|  !}}\exp\left(  -\frac{\tilde{\zeta}r^{2}}%
{2}\right)  \left(  \sqrt{\tilde{\zeta}}r\right)  ^{\left|  \alpha\right|
}e^{i\alpha\varphi-iEt/\hbar}.\label{50}%
\end{equation}
For this physical state, the corresponding $V\ ,V_{nc}$, $Q_{c},$ $Q_{nc}$ and
$\Sigma_{\theta}$ are
\begin{align}
V &  =\frac{1}{2}mw^{2}r^{2}\nonumber\\
V_{nc} &  =\left(  \frac{mw\theta}{2\hbar}\right)  ^{2}\frac{\alpha^{2}%
\hbar^{2}}{2mr^{2}}-\frac{m\theta w^{2}\alpha}{2}\nonumber\\
Q_{c} &  =-\frac{1}{2}mw^{2}r^{2}+\hbar w\left(  \left|  \alpha\right|
+1\right)  -\frac{\alpha^{2}\hbar^{2}}{2mr^{2}}\label{51}\\
Q_{nc} &  =\left[  \sqrt{1-\left(  \frac{mw\theta}{2\hbar}\right)  ^{2}%
}-1\right]  \hbar w\left(  \left|  \alpha\right|  +1\right)  -\left(
\frac{mw\theta}{2\hbar}\right)  ^{2}\frac{\alpha^{2}\hbar^{2}}{2mr^{2}%
}\nonumber\\
\Sigma_{\theta} &  =0.\nonumber
\end{align}

A first inspection on the potentials shows that in\ the lowest energy state,
characterized by $\alpha=0$ and $r=0,$ the zero point energy is given by
\begin{equation}
E_{0,0,\theta}=Q_{c}+Q_{nc}=\hbar w\left(  1-\left(  \frac{mw\theta}{2\hbar
}\right)  ^{2}\right)  ^{1/2},\label{52.5}%
\end{equation}
exactly what would be expected from quantum potentials, which must account for
all nonclassical behaviors.

According to the criteria proposed in the latter subsection, the condition for
the classical limit is $Q_{c}+Q_{nc}\rightarrow0$, while for the commutative
limit the condition is reduced to $V_{nc}+Q_{nc}\rightarrow0$. We start by the
commutative limit. The sum $V_{nc}+Q_{nc}$ can be written as
\begin{equation}
V_{nc}+Q_{nc}=-\left(  \frac{mw\theta}{2\hbar}\right)  \hbar w\alpha+\left[
\sqrt{1-\left(  \frac{mw\theta}{2\hbar}\right)  ^{2}}-1\right]  \hbar w\left(
\left|  \alpha\right|  +1\right)  .\label{52.7}%
\end{equation}
The sum of other terms in the Hamiltonian of (\ref{44.6}) is
\begin{equation}
\frac{\left(  \nabla S\right)  ^{2}}{2m}+V+Q_{c}=\hbar w\left(  \left|
\alpha\right|  +1\right) \label{52.8}%
\end{equation}
In order that (\ref{52.7})\ be negligible when compared to (\ref{52.8}), it is
sufficient that the parameters $m,w,\hbar\ $and $\theta$ satisfy
$mw\theta/2\hbar\ll1$, or in other words, that $\theta\ll2\hbar/mw.$ This is
the precise meaning that must be attributed to $\theta\rightarrow0$ in the
case of the noncommutative harmonic oscillator. Once this condition is
satisfied, all the observable quantities computed with the aid of the
oscillator wavefunction, as expected values and also the spectrum, will be
indistinguishable to the ones corresponding to its commutative counterpart.

There is a general comment that, what is important to recover the classical
world, is the smallness of $\hbar$ with respect to the physical quantities of
the problem\footnote{In the notation adopted in this work, such a claim is
written as $\hbar\rightarrow0$.}. As stated in when performing a theoretical
analysis in Section 3, and reinforced in Subsection 4.3, that statement is
ungrounded. By looking for the classical limit of the noncommutative harmonic
oscillator, it is easy to see that there is no parameter to vary in order that
the system assumes a classical behavior without canceling the interaction
term, which would be a trivial and uninteresting choice. Since $Q_{c}+Q_{nc}$
contains a term, $-mw^{2}r^{2}/2$, which do not depend on $\hbar,$ no matter
how small, but non-null, this constant is considered, the quantum effects
cannot be neglected. The classical limit of the harmonic oscillator in the
context of ordinary commutative quantum mechanics is only achieved by
constructing coherent states, which are superpositions of states containing
different quantum numbers \cite{38.5}. For the case of NCQM, however, that
procedure is non-trivial, and\ involves certain subtleties \cite{39.5}.

This example of application is useful to illustrate the spirit of the ideas
defended in this work, that the commutative and classical limits of a system
must be considered as realized on its physical states, rather on the equations
of motion.

\subsection{WKB Approximation in the Case of a One-Direction Interacting Potential}

Here, we present an example to complete the one of the latter subsection in a
discussion about noncommutative/commutative and quantum/classical passages. We
shall\ show how the commutative limit may be obtained in certain cases by
varying parameters other than $\theta.$

Consider a particle in two dimensional plane under the action of a
one-direction interacting potential. The noncommutative Schr\"{o}dinger
equation is given by
\begin{equation}
i\hbar\frac{\partial}{\partial t}\Psi=\left[  \frac{\hat{p}_{x}^{2}}{2m}%
+\frac{\hat{p}_{y}^{2}}{2m}+V\left(  x-\frac{\theta\hat{p}_{y}}{2\hbar
}\right)  \right]  \Psi.\label{59}%
\end{equation}
Writing the wavefunction as
\begin{equation}
\Psi\left(  x,y,t\right)  =e^{-iEt/\hbar+ik_{y}y}\psi(x),\label{60}%
\end{equation}
we separate the variables, obtaining
\begin{align}
-\frac{\hbar^{2}}{2m}\frac{d^{2}\psi}{dx^{2}}+V\left(  x-\frac{\theta k_{y}%
}{2}\right)  \psi &  =E_{x}\psi\label{61}\\
\frac{\hbar k_{y}}{2m} &  =E_{y}.\label{62}%
\end{align}

Performing a WKB approximation inserting the $ansatz$
\begin{equation}
\psi(x)=e^{i\phi(x)/\hbar},\text{ }\phi=\phi_{0}+\hbar\phi_{1}+...\label{63}%
\end{equation}
on (\ref{61}) we find, after simplification
\begin{equation}
\Psi_{WKB}^{\pm}(x,y,t)=\frac{C_{\pm}\exp\left[  \frac{i}{\hbar}\left(
\pm\int2m\left(  E_{x}-V-V_{nc}\right)  ^{1/2}dx-\left(  E_{x}+E_{y}\right)
t+\hbar k_{y}y\right)  \right]  }{\left[  2m\left(  E_{x}-V-V_{nc}\right)
\right]  ^{1/4}},\label{64}%
\end{equation}
where $C_{\pm}$ is a constant.

The conditions for the validity of this approximation are (see, for example,
\cite{18})
\begin{align}
m\hbar\left|  \left(  V+V_{nc}\right)  ^{\prime}\right|  /\left[  2m\left(
E_{x}-V-V_{nc}\right)  \right]  ^{3/2} &  \ll1,\nonumber\\
m\hbar^{2}\left|  \left(  V+V_{nc}\right)  ^{\prime\prime}\right|  /\left[
2m\left(  E_{x}-V-V_{nc}\right)  \right]  ^{2} &  \ll1,\nonumber\\
m\hbar^{3}\left|  \left(  V+V_{nc}\right)  ^{\prime\prime\prime}\right|  /
\left[  2m\left(  E_{x}-V-V_{nc}\right)  \right]  ^{5/2} &  \ll1,\label{65}%
\end{align}
and so on for higher derivatives of $\left(  V+V_{n}\right)  ^{\prime
}=d\left(  V+V_{nc}\right)  /dx$. The potentials corresponding to the
wavefunction (\ref{64}) are
\begin{align}
V_{nc} &  =V\left(  x-\frac{\theta k_{y}}{2}\right)  -V(x)\nonumber\\
Q_{c} &  =-\frac{\hbar^{2}}{2m}\left(  E_{x}-V\right)  ^{1/4}\frac{d^{2}%
}{dx^{2}}\left(  E_{x}-V\right)  ^{-1/4}\nonumber\\
Q_{nc} &  =-\frac{\hbar^{2}}{2m}\left(  E_{x}-V-V_{nc}\right)  ^{1/4}%
\frac{d^{2}}{dx^{2}}\left(  E_{x}-V-V_{nc}\right)  ^{-1/4}+\frac{\hbar^{2}%
}{2m}\left(  E_{x}-V\right)  ^{1/4}\frac{d^{2}}{dx^{2}}\left(  E_{x}-V\right)
^{-1/4}\label{66}\\
\Sigma_{\theta} &  =0\nonumber
\end{align}

Notice that $\theta$ appears multiplying$\ k_{y}=k\sin\varphi$, where
$\varphi$ is the angle between the direction of motion and the direction where
the potential acts. It can thus be chosen as our parameter to vary in order to
measure the degree of deviation from the commutative behavior. Let us suppose
that $\theta$ is sufficiently small (as it should be)\ in order that
(\ref{29.5}) be undetectable, but ``big'' enough to have some influence on
other detectable quantities, such as energy spectrum, etc. If $\varphi$ is
small and renders $V_{nc}+Q_{nc}$ negligible as compared to the other terms of
the Hamiltonian in the Hamilton-Jacobi equation,\ the system assumes a
commutative behavior. When the speed of the particle is sufficiently slow,
this limit is achieved for relatively\ large angles.

From the conditions (\ref{65}), one can infer that $\left|  Q_{c}%
+Q_{nc}\right|  \ll E_{x}-V-V_{nc}$, and thus the wavefunction is
semiclassical if they are satisfied. Observe that the turning points, where
$E_{x}-V-V_{nc}=0$ and the approximation fails, are shifted in comparison with
the ones of the analogous commutative problem due to the presence of $V_{nc}.$

\section{Discussion and Outlook}

In this work, we presented an alternative interpretation for the meaning of
the canonical noncommutativity obtained from arguments of string theory. In
order to justify our point of view on the validity of interpreting (\ref{1})
as being realized just by the coordinate observables of the particles, and not
by the spacetime itself, we discussed some of the aspects of the derivation of
the noncommutativity in the Landau problem and in the string context.
Arguments for the existence of a loophole in the usual interpretation of the
results found in the literature were presented.

When we consider the stringy noncommutativity from the new point of view, the
interpretation of the Weyl symbols as points of the spacetime, considered as
senseless in \cite{13,27,27.5}, but adopted by the great majority of the
string inspired articles, like \cite{7}, for example, is perfectly justified.
Although these articles do not presuppose our interpretation for
noncommutativity, their calculations\ are\ in accordance with it. The possible
problem in the legitimacy of the use of the Feynman graph approach on
noncommutative spaces, which was underlined in a discussion carried on in
\cite{15}, is absent here, since NCQFT is assumed as being based on a
commutative spacetime.

In the context of quantum mechanics, the implications of adopting\ the new
point of view for noncommutativity are manifest into the foundations. The
common assertion that the concept wavefunction does not make sense in
noncommutative spaces \cite{13,27.5} looses its validity for the stringy
noncommutativity considered because it is not of the spacetime coordinates.
The wavefunction, however, has a different meaning in NCQM than in its
commutative counterpart. It is possible to define a consistent probability
density given by the expression $\rho=\left|  \Psi\right|  ^{2}$, but the
interpretation of $\rho(\vec{x},t)d^{3}x$ is that of being the probability of
finding the canonical coordinate of the particle in the volume $d^{3}x$ around
the point $\vec{x}$ at time $t$.

Other related topic of interest discussed was the use of the Schr\"{o}dinger
formulation and the associated Hamilton-Jacobi formalism to perform an
analysis of the structure of NCQFT and NCQM and their possible limiting
behaviors. A formal analysis, comparing the potential terms according to the
derivatives present in their definitions, revealed\ that\ NCQM is richer than
NCQFT from this point of view. While the\ complete description of\ NCQFT\ can
be done by considering the properties of the\ three potentials $V, $ $V_{nc}$
and$\ Q_{K}$, the corresponding one for NCQM is more complex, needing the four
potentials $V,$ $V_{nc}$, $Q_{K},$ $Q_{I}$ and the $\Sigma_{\theta}$ term
present in the probability conservation equation.

The conditions for the attainment of the classical and commutative limits of
NCQFT were determined as being $Q_{K}=Q_{c}+Q_{nc}\rightarrow0$ and
$V_{nc}+Q_{nc}\rightarrow0$, respectively. For NCQM, our considerations were
restricted to the case where $\Sigma_{\theta}\rightarrow0$, which was the one
of interest for the examples of application presented. The conditions for the
classical limit were found to be $Q_{K}+Q_{I}=Q_{c}+Q_{nc}\rightarrow0$, and
for the commutative one $V_{nc}+Q_{nc}\rightarrow0$.

In the same way that $\hbar$ must not be considered as the agent responsible
for the quantum effects in ordinary and noncommutative field theories and
quantum mechanics, $\theta^{ij}$ must not be considered as the one responsible
for the noncommutative manifestations in NCQFT and NCQM. Although achieving
the commutative limit by setting $\theta^{ij}\rightarrow0$ was possible in the
examples presented, this route for the classical limit will not always be
available when Hamiltonians\ containing more complicated potentials are
considered \cite{40}. In these cases the formalism of this work may be useful
in finding other routes for the attainment of the commutative behavior.

There is the immediate possibility to apply the Hamilton-Jacobi formalism
present in this work\ to investigate the properties\ of some simple models of
noncommutative\ scalar field theory or\ noncommutative quantum mechanical
systems whose Hamiltonians possesses higher-order derivatives or are nonlocal,
like, for example, the anharmonic oscillator and the hydrogen atom. The
extension of the ideas presented here is in principle possible for other
theories than the scalar field ones, like the noncommutative version of\ QED.

\section*{Acknowledgments}

The author is greatly indebted to Nelson Pinto Neto for valuable discussions,
suggestions and for all the\ corrections\ on an earlier manuscript. We also
acknowledge Ronaldo Penna Neves for discussions in the earlier stage of this
work, and Jos\'{e} Hel\"{a}yel Neto for discussions, encouragement and
suggestions on this manuscript. This work was financially supported by CAPES.

\end{document}